\documentclass[fleqn,12pt]{article}

\usepackage{arxiv}

\usepackage{amsmath}
\usepackage{amssymb}
\usepackage{enumerate}
\usepackage{xcolor}
\usepackage{hyperref}
\usepackage{pgfplots}
\usepackage{float}
\usepackage{libertine}
\usepackage{mathtools}
\usepackage{subfig}
\usepackage[squaren,Gray]{SIunits}
\usepackage{tabto}

\usepackage{empheq}
\usepackage{algorithm}
\usepackage{algpseudocode}

\newtheorem{theorem}{Theorem}
\newtheorem{lemma}[theorem]{Lemma}
\newtheorem{property}[theorem]{Property}
\usepackage{enumitem}
\usepackage{array}
\usepackage[square,sort,comma,numbers]{natbib}
\usepackage{lipsum}
\usepackage{graphicx}
\usepackage{multirow}
\usepackage{color}
\usepackage{amsmath}
\usepackage{amssymb}
\usepackage[export]{adjustbox}
\usepackage{bbm}
\usepackage{subfig}
\usepackage{booktabs}

\setlist[itemize]{leftmargin=*,nolistsep,noitemsep}

\usepackage{tikz, tikz-3dplot}
\usepackage{pgfplots} 
\usepgfplotslibrary{patchplots}
\pgfplotsset{compat=newest}
\usetikzlibrary{decorations.pathreplacing}
\usetikzlibrary{fadings}
\usetikzlibrary{positioning, fit, arrows.meta, shapes,calc,chains,scopes}
\usetikzlibrary{matrix,arrows,backgrounds}
\usetikzlibrary{tikzmark}
\usepackage{amsmath,amssymb}
\usetikzlibrary{shapes.multipart}
\tikzset{elementwiseoperation/.style={circle, draw=green!55!blue, fill=green!55!blue, inner sep=0pt},
    elementwisefunction/.style={ellipse, draw=green!55!blue, fill=green!55!blue, inner sep=1pt},
    ct/.style={rectangle,, draw, minimum width=1cm, inner sep=1pt},
    sqt/.style={rectangle, draw=green!55!blue,fill=green!55!blue, minimum width=1cm, minimum height=1cm},
    sqt2/.style={circle, draw, fill=blue, minimum width=2cm, minimum height=2cm},
    gt/.style={rectangle, draw, minimum width=4mm, minimum height=3mm, inner sep=1pt},
    mylabel/.style={font=\scriptsize\sffamily},
    neuron/.style={ 
    circle,draw,thick, 
    inner sep=0pt, 
    minimum size=2.5em, 
    node distance=1ex and 2em, 
    fill=green!55!blue,
  },
  group/.style={ 
    rectangle,draw,thick, 
    inner sep=0pt, 
  },
  io/.style={ 
    neuron, 
    fill=gray!15, 
  },
  conn/.style={ 
    -{Straight Barb[angle=60:2pt 3]}, 
    thick, 
  },}%

\usepackage[utf8]{inputenc}
\usepackage{libertine}
\usepackage{graphicx}
\usepackage{color}
\usepackage{relsize}
\usepackage{apptools}
\usepackage{ragged2e}
\usepackage{multirow}
\usepackage{multicol}
\usepackage{empheq}
\usepackage[most]{tcolorbox}
\usepackage{enumitem}

\usepackage{supertabular}
\usepackage{animate}
\usepackage{media9}
\usepackage{tabto}
\usepackage[page,toc,title]{appendix} 
\usepackage{amsmath,amssymb}
\usepackage{xcolor}

\usepackage[intoc]{nomencl}
\makenomenclature

\PassOptionsToPackage{hyphens}{url}\usepackage{hyperref}
\usepackage{array}
\newcolumntype{L}[1]{>{\flushleft\arraybackslash}p{#1}}

\usepackage{tikz, tikz-3dplot}
\usepackage{pgfplots} 
\usepgfplotslibrary{patchplots}
\pgfplotsset{compat=newest}
\usetikzlibrary{decorations.pathreplacing}
\usetikzlibrary{fadings}
\usetikzlibrary{positioning, fit, arrows.meta, shapes,calc,chains,scopes}
\usetikzlibrary{matrix,arrows,backgrounds}
\usetikzlibrary{tikzmark}
\usepackage{mathptmx}
\usepackage{calc}\usetikzlibrary{tikzmark,calc,,arrows,shapes,decorations.pathreplacing}
\tikzset{every picture/.style={remember picture}}
\usetikzlibrary{shapes.multipart}
\tikzset{elementwiseoperation/.style={circle, draw=green!55!blue, fill=green!55!blue, inner sep=0pt},
    elementwisefunction/.style={ellipse, draw=green!55!blue, fill=green!55!blue, inner sep=1pt},
    ct/.style={rectangle,, draw, minimum width=1cm, inner sep=1pt},
    sqt/.style={rectangle, draw=green!55!blue,fill=green!55!blue, minimum width=1cm, minimum height=1cm},
    sqt2/.style={circle, draw, fill=blue, minimum width=2cm, minimum height=2cm},
    gt/.style={rectangle, draw, minimum width=4mm, minimum height=3mm, inner sep=1pt},
    mylabel/.style={font=\scriptsize\sffamily},
    neuron/.style={ 
    circle,draw,thick, 
    inner sep=0pt, 
    minimum size=2.5em, 
    node distance=1ex and 2em, 
    fill=gray,
  },
  group/.style={ 
    rectangle,draw,thick, 
    inner sep=0pt, 
  },
  io/.style={ 
    neuron, 
    fill=gray!15, 
  },
  conn/.style={ 
    -{Straight Barb[angle=60:2pt 3]}, 
    thick, 
  },}%
\usepackage{esvect}

\newcommand{\trsign}{{\mathsf{T}}}
\newcommand{\tr}{^{\trsign}}

\makeatletter
\newsavebox\zb@x
\newcounter{z@@m}
\usepackage{calc}
\newdimen\B@r\newdimen\P@r
\newdimen\@zw\newdimen\@zh\newdimen\@zd

\newcommand{\zoombox}[2][0]{%
  \leavevmode%
  \sbox\zb@x{#2}%
  \setlength\B@r{1pt*\ratio{\wd\zb@x}{\ht\zb@x+\dp\zb@x}}%
  \setlength\P@r{1pt*\ratio{\paperwidth}{\paperheight}}%
  \ifdim\B@r>\P@r\relax%
    \setlength\@zw{\wd\zb@x}\setlength\@zh{\@zw*\ratio{\paperheight}{\paperwidth}}%
    \setlength\@zd{(\@zh-\ht\zb@x-\dp\zb@x)*\real{0.5}+\dp\zb@x}%
    \setlength\@zh{\@zh-\@zd}%
  \else%
    \setlength\@zh{\ht\zb@x+\dp\zb@x}%
    \setlength\@zw{\@zh*\ratio{\paperwidth}{\paperheight}}%
    \setlength\@zh{\ht\zb@x}\setlength\@zd{\dp\zb@x}%
  \fi%
  \makebox[0pt][l]{\makebox[\wd\zb@x][c]{\makebox[\@zw][l]{%
    \pdfdest name {zbfs\thez@@m} fitr
      width  \@zw\space
      height \@zh\space
      depth  \@zd\space
  }}}%
  \pdfdest name {zb\thez@@m} fitr
    width  \wd\zb@x\space
    height \ht\zb@x\space
    depth  \dp\zb@x\space
  \immediate\pdfannot 
    width  \wd\zb@x\space
    height \ht\zb@x\space
    depth  \dp\zb@x\space
  {%
    /Subtype/Link/H/N
    /Border [0 0 #1 [1 2]]
    /A <<
      /S/JavaScript
      /JS (
        if(typeof(zoomed)=='undefined'||!zoomed){
          var lastView=this.viewState;
          if(app.fs.isFullScreen) this.gotoNamedDest('zbfs\thez@@m');
          else this.gotoNamedDest('zb\thez@@m');
          zoomed=true;
        }else{
          this.viewState=lastView;
          zoomed=false;
        }
      )
    >>
  }%
  \usebox{\zb@x}%
  \stepcounter{z@@m}%
} 
\makeatother

\graphicspath{{figs/}}

\title{Learning  Neural Optimal Interpolation Models and Solvers}

\author{
Maxime Beauchamp\\
IMT Atlantique\\
maxime.beauchamp@imt-atlantique.fr \\
\And
Joseph Thompson\\
IMT Atlantique\\ 
joseph.thompson@imt-atlantique.fr
\And
Hugo Georgenthum\\
IMT Atlantique\\ 
hugo.georgenthum@imt-atlantique.fr
\And
Quentin Febvre\\
IMT Atlantique\\ 
quentin.febvre@imt-atlantique.fr
\And
Ronan Fablet\\
IMT Atlantique\\ 
ronan.fablet@imt-atlantique.fr
}

\begin{document}

\maketitle

\begin{abstract}
The reconstruction of gap-free signals from observation data is a critical challenge for numerous application domains, such as geoscience and space-based earth observation, when the available sensors or the data collection processes lead to irregularly-sampled and noisy observations. Optimal interpolation (OI), also referred to as kriging, provides a theoretical framework to solve interpolation problems for Gaussian processes (GP). The associated computational complexity being rapidly intractable for n-dimensional tensors and increasing numbers of observations, a rich literature has emerged to address this issue using ensemble methods, sparse schemes or iterative approaches. Here, we introduce a neural OI scheme. It exploits a variational formulation with convolutional auto-encoders and a trainable iterative gradient-based solver. Theoretically equivalent to the OI formulation, the trainable solver asymptotically converges to the OI solution when dealing with both stationary and non-stationary linear spatio-temporal GPs. Through a bi-level optimization formulation, we relate the learning step and the selection of the training loss to the theoretical properties of the OI, which is an unbiased  estimator with minimal error variance. Numerical experiments for 2{\sc d}+t synthetic GP datasets demonstrate the relevance of the proposed scheme to learn computationally-efficient and scalable OI models and solvers from data. As illustrated for a real-world interpolation problems for satellite-derived geophysical dynamics, the proposed framework also extends to non-linear and multimodal interpolation problems and significantly outperforms state-of-the-art interpolation methods, when dealing with very high missing data rates. 
\end{abstract}
\section{Introduction}
\label{sec:intro}

Interpolation problems are critical challenges when dealing with irregularly-sampled observations. Among others, Space earth observation, geoscience, ecology, fisheries generally monitor a process of interest through gappy observations due to the characteristics of the sensors and/or the data collection process. As illustrated in Fig.\ref{xp_SSH} for satellite-based earth observation, missing data rates may be greater than 90\%, which makes the interpolation problem highly challenging.

Optimal Interpolation (OI) \cite{cressie_statistics_2015}, also referred to as kriging \citep{chiles_2012}, provides a theoretical framework to address such interpolation problems for Gaussian processes. Given the covariance structure of the process of interest along with the covariance of the observation noise, one can derive the analytical OI solution. For high-dimensional states, such as space-time processes, the computation of this analytical solution rapidly becomes intractable as it involves the inversion of a $N\times N$ matrix with $N$ the number of observation points. When dealing with space-time processes, OI also relates to data assimilation \cite{asch_data_2016,evensen_2022}. In this context, Kalman methods, including ensemble-based extensions, exploit the sequential nature of the problem to solve the OI problem allowing for dynamical flow propagation of the uncertainties. Overall, this broad category of geostatistical approaches is the state-of-the-art in numerous domains, such as geoscience, climate science, ecology\ldots 

Data-driven and learning-based approaches have also received a growing interest to address interpolation problems \citep{barth_2022, beauchamp_2020, lguensat_data-driven_2017}, while Image and video inpainting are popular interpolation problems in computer vision \citep{huang_2016, kim_2019, lethuc_2017, xu_2019}. As they typically relate to object removal applications or restoration problems, they usually involve much lower missing data rates than the ones to be delt with in the above-mentionned applications and address natural images, which are likely not representative of space-time dynamics addressed in geoscience, meteorology, ecology\ldots A recent literature has also emerged to exploit deep learning methods to solve inverse problems classically stated as the minimization of a variational cost. This includes neural architectures based on the unrolling of minimization algorithms \cite{mccann_convolutional_2017,lucas_using_2018,wei_tuning-free_2020}.    

Here, we introduce a neural OI framework. Inspired by the neural method introduced in \cite{Fablet_2021} for data assimilation, we develop a variational formulation based on convolutional auto-encoders and introduce an associated trainable iterative gradient-based solver. Our key contributions are four-fold:
\begin{itemize}
    \item We show that our variational formulation is equivalent to OI when dealing with Gaussian processes driven by linear dynamics. Under these assumptions, our trainable iterative gradient-based solver converges asymptotically towards the OI solution;
    \item Regarding the definition of the training losses, we relate the learning step of the proposed neural architecture to the properties of the OI solution as an unbiased estimator with minimal error variance;
    \item Our framework extends to learning optimal interpolation models and solvers for non-linear/non-Gaussian processes and multimodal observation data;
    \item Numerical experiments for a 2{\sc d}+t Gaussian process support the theoretical equivalence between OI and our neural scheme for linear Gaussian case-studies. They also illustrate the targeted scalable acceleration of the interpolation. 
    \item We report a real-world application to the interpolation of sea surface dynamics from satellite-derived observations. Our neural OI scheme significantly outperforms the state-of-the-art methods and can benefit from multimodal observation data to further improve the reconstruction performance.
\end{itemize}
To make easier the reproduction of our results, an open-source version of our code is available \footnote{To be made available in a final version}.

This paper is organized as follows. Section \ref{sec: problem statement} formally introduces optimal interpolation and related work. We present the proposed neural OI framework in Section \ref{sec: 4dvarnet-oi}. Section \ref{sec: experiments} reports numerical experiments for both synthetic GP datasets and real-world altimetric sea surface observations. We discuss further our main contributions in Section \ref{sec: conclusion}.

\section{Problem statement and Related work}
\label{sec: problem statement}

For a n-dimensional Gaussian process $\mathbf{x}$ with mean $\mu$ and covariance $\mathbf{P}$, the optimal interpolation states the reconstruction of state $\mathbf{x}$ from noisy and gappy observations $\mathbf{y}$ as the minimization of a variational cost:
\begin{equation}
\label{eq: OI}
\widehat{\mathbf{x}}=\arg \min_\mathbf{x}  \left \|\mathbf{y}-H_\Omega \cdot \mathbf{x}  \right \|^2_{\mathbf{R}}+ \left \|\mathbf{x}-\mu \right \|^2_{\mathbf{P}}
\end{equation}
with $H_\Omega$ denotes the observation matrix to map state $\mathbf{x}$ over domain $\cal{D}$ to the observed domain $\Omega$. $\|\cdot\|^2_{\mathbf{R}}$ is the Mahanalobis norm w.r.t. the covariance of the observation noise $\mathbf{R}$ and $\|\cdot \|^2_\mathbf{P}$ the Mahanalobis distance with {\em a priori} covariance  $\mathbf{P}$. The latter decomposes as a 2-by-2 block matrix $[\mathbf{P}_{\Omega,\Omega} \mathbf{P}_{\Omega,\overline{\Omega}};\mathbf{P}_{\overline{\Omega},\overline{\Omega}} \mathbf{P}^{T}_{\Omega,\overline{\Omega}}]$ with $\mathbf{P}_{\cal{A},\cal{A}'}$ the covariance between subdomains $\cal{A}$ and $\cal{A}$ of domain $\cal{D}$.

The OI variational cost (\ref{eq: OI}) being linear quadratic, the solution of the optimal interpolation problem is given by:
\begin{equation}
\label{eq: OI solution}
\widehat{ \mathbf{x}}= \mu + \mathbf{K} \cdot \mathbf{y}
\end{equation}
with $\mathbf{K}$ referred to as the Kalman gain $ \mathbf{P} H\tr_\Omega ( H_\Omega \mathbf{P} H\tr_\Omega + \mathbf{R} )^{-1}$, where $\mathbf{P} H\tr_\Omega$ is the (grid,obs) prior covariance matrix, $H_\Omega \mathbf{P} H\tr_\Omega$ is the (obs,obs) prior covariance matrix. For high-dimensional states, such as n{\sc d} and  n{\sc d}$+t$ states, and large observation domains, the computation of the Kalman becomes rapidly intractable due to the inversion of a $|\Omega|\times|\Omega|$ covariance matrix. This has led to a rich literature to solve minimization \ref{eq: OI} without requiring the above-mentioned $|\Omega|\times|\Omega|$ matrix inversion, among others gradient-based solvers using matrix-vector multiplication (MVMs) reformulation 
\cite{Pleiss_2020, Aune_2013, Charlier_2020, Cutajar_2016},  methods based on sparse matrix decomposition with tapering \citep{furrer_2006,romary_2018} or precision-based matrix parameterizations 
\citep{lindgren_2011, carrizovergara_2018, clarotto_2022}.\\

Variational formulations have also been widely explored to solve inverse problems. Similarly to (\ref{eq: OI}), the general formulation involves the sum of a data fidelity term and of a prior term \citep{asch_data_2016}. In a model-driven approach, the latter derives from the governing equations of the considered processes. For instance, data assimilation in geoscience generally exploits PDE-based terms to state the prior on some hidden dynamics from observations. In signal processing and computational imaging, similar formulations cover a wide range of inverse problems, including inpainting issues \citep{bertalmio_navier-stokes_2001}:
\begin{align}
\label{eq: 4dvar}
\widehat{\mathbf{x}}&=\arg \min_\mathbf{x} \mathcal{J}_\Phi (\mathbf{x},\mathbf{y},\boldsymbol{\Omega}) \nonumber \\
                    &=\mathcal{J}^o(\mathbf{x},\mathbf{y}) + \lambda \left \| \mathbf{x}-\Phi (\mathbf{x}) \right \|^2  
\end{align}
$\mathcal{J}^o(\mathbf{x},\mathbf{y})$ is the data fidelity term which is problem-dependent. The prior regularization term $\| \mathbf{x}-\Phi (\mathbf{x}) \|^2$ can be regarded as a projection operator. This parameterization of the prior comprises both gradient-based priors using finite-difference approximations, proximal operators as well as plug-and-play priors \cite{aubert_mathematical_2006,lucas_using_2018,mccann_convolutional_2017}. As mentioned above, these formualtions have also gained interest in the deep learning literature for the definition of deep learning schemes based on the unrolling of minimization algorithms \citep{andrychowicz2016learning,aggarwal_modl_2019} for (\ref{eq: 4dvar}).  

Here, we further explore the latter category of approaches to solve optimal problems stated as (\ref{eq: OI}), including when covariance $\mathbf{P}$ is not known a priori.

\section{Neural OI framework}
\label{sec: 4dvarnet-oi}

This Section presents the proposed trainable OI framework. We first introduce the proposed neural OI solver (Section \ref{sec: OI solver}) and the associated learning setting (Section \ref{sec: learning}). We then describe extensions to non-linear and multimodal interpolation problems.

\subsection{Neural OI model and solver}
\label{sec: OI solver}

Let us introduce the following variational formulation to reconstruct state $\mathbf{x}$ from gappy observations $\mathbf{y}$:
\begin{equation}
\label{eq: proj-based OI}
\widehat{\mathbf{x}}=\arg \min_\mathbf{x} \left \|\mathbf{y}-H_\Omega \cdot \mathbf{x}\right \|^2+ \lambda \left \| \mathbf{x}-\Phi (\mathbf{x}) \right \|^2  
\end{equation}
where $\lambda$ a positive scalar to balance the data fidelity term and the prior. $\Phi(\cdot)$ is a linear neural auto-encoder which states the prior onto the solution. Hereafter, we denote $\mathcal{J}_\Phi ( \mathbf{x}, \mathbf{y}, \Omega ) = \|\mathbf{x}-\mathbf{y} \|^2 + \lambda \| \mathbf{x}-\Phi (\mathbf{x}) \|^2$. 

\begin{property}  Variational formulation (\ref{eq: proj-based OI}) is equivalent to optimal interpolation problem (\ref{eq: OI})  when considering a matrix parameterization of the prior $\Phi(x) = (\mathbf{I}-\mathbf{L})\mathbf{x}$ with $\mathbf{L}$ the square-root (as a Cholesky decomposition) of $\mathbf{P}$ and a spherical observation covariance, {\em i.e.} $\mathbf{R} = \sigma^2 \mathbbm{1}$.
\end{property}
The proof comes immediately when noting that the regularization term of the variational cost also writes:
  $\mathbf{x}\tr\mathbf{P}^{-1}\mathbf{x} = \mathbf{x}\tr\mathbf{L}\tr\mathbf{L}\mathbf{x} = ||\mathbf{L}\mathbf{x}||^2 = ||\mathbf{x}-\Phi\mathbf{x}||^2 $.

\begin{lemma} 
\label{lemma: Oi equivalence}
For a stationary Gaussian process and a Gaussian observation noise with $\mathbf{R} = \sigma^2 \mathbf{I}$, we can restate the associated optimal interpolation problem (\ref{eq: OI}) as minimization problem (\ref{eq: proj-based OI}) with neural operator $\Phi(\cdot)$ being a linear convolutional network.
\end{lemma}
The proof results from the translation-invariant property of the covariance of stationary Gaussian processes. Computationally, we can derive $\Phi$ as the inverse Fourier transform of the square-root of the Fourier transform of covariance $\mathbf{P}$ in (\ref{eq: OI}) as exploited in Gaussian texture synthesis \cite{galerne_random_2011}.

Lemma \ref{lemma: Oi equivalence} provides the basis to learn a solver of variational formulation (\ref{eq: proj-based OI}) to address optimal interpolation problem (\ref{eq: OI}). We benefit from automatic differentiation tools associated with neural operators to investigate iterative gradient-based solvers as introduced in meta-learning \citep{andrychowicz2016learning, hospedales_meta-learning_2020}, see Algorithm \label{alg:gradient solver}. The latter relies on an iterative gradient-based update where neural operator ${\mathcal{G}}$ combines a LSTM cell \citep{Shi_2015} and a linear layer to map the hidden state of the LSTM cell to the space spanned by state $\mathbf{x}$. Through the LSTM may capture long-term dependencies, operator ${\mathcal{G}}$ defines a momentum-based gradient descent. Overall, this class of generic learning-based methods was explored and referred to as 4DVarNet schemes in \cite{fablet2020joint} for data assimilation problems. Here, as stated in Lemma \ref{lemma: grad descent}, we parameterize weighting factors $a(\cdot)$ and $\omega(\cdot)$ such that the LSTM-based contribution dominates for the first iterations while for a greater number of iterations the iterative update reduces to a simple gradient descent. Hereafter, we refer to the proposed neural OI framework as 4DVarNet-OI.

\begin{algorithm}
\caption{Iterative gradient-based solver for (\ref{eq: proj-based OI}) given initial condition $\mathbf{x}^{(0)}$, observation $\mathbf{y}$ and sampled domain  $\Omega$. Let $a(\cdot)$ and $\omega(\cdot)$ be positive scalar functions and ${\mathcal{G}}$ a LSTM-based neural operator.  }\label{alg:gradient solver}
\begin{algorithmic}
\State $\mathbf{x} \gets \mathbf{x}^{(0)}$
\State $k \gets 0$
\While{$k \leq K$}
\State $k \gets k+1$
\State $\mathbf{g} \gets \nabla_\mathbf{x} \mathcal{J}_\Phi\left ( \mathbf{x}^{(k)} , \mathbf{y}, \Omega\right )$
\State $ \mathbf{x} \gets \mathbf{x} - a(k) \cdot \left [ \omega(k) \cdot \mathbf{g} + \left (1-\omega(k) \right )  \cdot \mathcal{G}\left( \mathbf{g} \right ) \right ]$
\EndWhile
\end{algorithmic}
\end{algorithm}

\begin{lemma} 
\label{lemma: grad descent}
Let us consider the following parameterizations for functions $a(\cdot)$ and $\omega(\cdot)$
\begin{align}
a(k)=  \frac{\nu \cdot K_0}{K_0+k} \mbox{  ;  } 
\omega(k)= \tanh \left ( \alpha \cdot (k -K_1 )  \right )
\end{align}
where $\nu$ and $\alpha$ are positive scalars, and $K_{1,2}$ positive integers. 
If ${\mathcal{G}}(\cdot)$ is a bounded operator and $\Phi(\cdot)$ is a linear operator given by $\mathbf{I}  - \mathbf{P}^{1/2}$, then Algorithm \ref{alg:gradient solver} converges towards the solution (\ref{eq: OI solution}) of the minimization of optimal interpolation cost (\ref{eq: proj-based OI}).
\end{lemma}
The proof of this lemma derives as follows. As ${\mathcal{G}}(\cdot)$ is bounded, the considered parameterization of the gradient step in Algorithm \ref{alg:gradient solver} is asymptotically equivalent to a simple gradient descent with a decreasing step size. Therefore, it satisfies the convergence conditions towards the global minimum for a linear-quadratic variational cost \cite{borkar_stochastic_2008}. We may highlight that the same applies with a stochastic version of Algorithm \ref{alg:gradient solver} and a convex variational cost \cite{borkar_stochastic_2008}.

The boundedness of operator $\mathcal{G}$ derives from that of the LSTM cell. Therefore, Lemma \ref{lemma: grad descent} guarantees that Algorithm \ref{alg:gradient solver} with a LSTM-based parameterization for operator $\mathcal{G}$ converges to the minimum of optimal interpolation cost (\ref{eq: proj-based OI}) whatever the parameters of $\mathcal{G}$. In this setting, operator $\mathcal{G}$ aims at accelerating the convergence rate of the gradient descent towards analytical solution (\ref{eq: OI solution}). Overall, we define  $\Theta=\{\Phi,\mathcal{G}\}$ and $\Psi^K_{\Theta} ( \mathbf{x}^\star , \mathbf{y} , \Omega )$ the interpolated state resulting from the application of Algorithm \ref{alg:gradient solver} with $K$ iterations from initial condition $\mathbf{x}^\star$ given observation data $\{\mathbf{y} , \Omega\}$. 

\subsection{Learning setting}
\label{sec: learning}

Formally, we state the training of the considered neural OI scheme (\ref{eq: proj-based OI}) according to a bi-level optimization problem
\begin{align}
\label{eq: bi-level}
\widehat{\Theta}= & \arg \min_{\Theta} {\cal{L}} \left ( \{\mathbf{x}_k, \mathbf{y}_k,\Omega_k,\widehat{\mathbf{x}}_k\} \right )  \nonumber \\ 
&\mbox{  s.t.  } 
\widehat{\mathbf{x}}_k = \arg \min_{\mathbf{x}_k} \mathcal{J}_\Phi \left ( \mathbf{x}_k, \mathbf{y}_k, \Omega_k \right ) 
\end{align}
where $\{\mathbf{x}_k,\mathbf{y}_k,\Omega_k\}$, where $k$ denotes the time index along the data assimilation window (DAW) $[t-k\Delta t;t+k\Delta t]$, is a training dataset with true states $\{\mathbf{x}_k\}$, gappy and noisy observations $\{\mathbf{y}_k\}$ and observation domains $\{\Omega_k\}$. \\
Overall, ${\cal{L}} ( \{\mathbf{x}_k, \mathbf{y}_k,\widehat{\mathbf{x}}_k\} )$ defines a training loss. 

\begin{lemma}
\label{lemma: bi-level}
Let us consider Optimal Interpolation problem (\ref{eq: OI}) with a spherical observation covariance $\mathbf{R}=\sigma^2\cdot\mathbf{I}$ and prior covariance $\mathbf{P}$. Let us parameterize trainable operator $\Phi$ in (\ref{eq: proj-based OI}) as a linear convolution operator. Optimal interpolation (\ref{eq: OI solution}) is then solution of a bi-level optimization problem (\ref{eq: bi-level}) with $\Phi = \mathbf{I}-\mathbf{P}^{1/2}$ for each of the following training losses:
\begin{align}
{\cal{L}}_1 \left ( \{\mathbf{x}_k, \mathbf{y}_k,\widehat{\mathbf{x}}_k\} \right ) &= \sum_k \| \mathbf{x}_k - \widehat{\mathbf{x}}_k\|^2 \nonumber \\
{\cal{L}}_2 \left ( \{\mathbf{x}_k, \mathbf{y}_k,\widehat{\mathbf{x}}_k\} \right ) &= \sum_k \| \mathbf{x}^{OI}_k - \widehat{\mathbf{x}}_k\|^2 \nonumber \\
{\cal{L}}_3 \left ( \{\mathbf{x}_k, \mathbf{y}_k,\widehat{\mathbf{x}}_k\} \right ) &= \sum_i 
\left \|\mathbf{y}_k-H_\Omega \cdot \mathbf{x}_k \right \|^2_{\mathbf{R}}+ \left \|\mathbf{x}_k \right \|^2_{\mathbf{P}}
\hspace*{-0.25cm}
\end{align}
where ${\cal{L}}_1$, ${\cal{L}}_2$ respectively denotes the mean squared error (MSE) w.r.t true states, MSE w.r.t OI solution denoted as $\mathbf{x}^{OI}$ and given by (\ref{eq: OI solution}) for covariance $(\mathbf{I}-\Phi)^2$. ${\cal{L}}_3$ stands for the OI variational cost.
\end{lemma} 
The proof results from the equivalence between variational formulations (\ref{eq: OI}) and (\ref{eq: proj-based OI}) under parameterization $\Phi = \mathbf{I}-\mathbf{P}^{1/2}$ and the property that OI solution (\ref{eq: OI solution}) is a minimum-variance unbiased estimator \cite{cressie_statistics_2015}.

This lemma motivates the following training setting for the proposed scheme:
\begin{align}
\label{eq: training loss}
\widehat{\Theta}=&\arg \min_{\Theta} {\cal{L}} \left ( \{\mathbf{x}_k,\mathbf{y}_k,\widehat{\mathbf{x}_k} \} \right ) \nonumber \\
& \mbox{  s.t.  } 
\widehat{\mathbf{x}}_k = \Psi^K_{\Theta} \left ( \{ \mathbf{x}^\star_k , \mathbf{x}_k , \Omega_k \}\right )
\end{align}
where ${\cal{L}}$ is among the training losses introduced in Lemma \ref{lemma: bi-level} and $\mathbf{x}^\star_k$ is an initial condition. ${\cal{L}}_{2,3}$ both require the explicit definition and parameterization of prior covariance $\mathbf{P}$ and ${\cal{L}}_{2}$ imposes to compute the OI analytical solution (\ref{eq: OI solution}) for the training dataset. In such situations, the proposed training framework aims at delivering a fast and scalable computation of (\ref{eq: OI solution}). By contrast, loss ${\cal{L}}_1$ only relies on the true states with no additional hypothesis on the underlying covariance, which makes it more appealing for most applications, see the experimental conclusions in Section \ref{sec: experiments}.

To train jointly solver component $\mathcal{G}$ and operator $\Phi$, we vary initial condition $\mathbf{x^\star_i}$ between some initialization $\mathbf{x}^{(0)}_i$ and detached outputs of Algorithm \ref{alg:gradient solver} for a predefined number of total iteration steps. This strategy also provides a practical solution to the memory requirement, which rapidly increases with the number of iterations during the training phase due to the resulting depth of the computational graph. In all the reported experiments, we use Adam optimizer over 200 epochs.


\subsection{Extension to non-linear and multimodal optimal interpolation}
\label{subsec: mutimodal OI}

While the analytical derivation of solution (\ref{eq: OI solution}) requires to consider a linear-quadratic formulation in both (\ref{eq: OI}) and (\ref{eq: proj-based OI}), Algorithm \ref{alg:gradient solver} applies to any differentiable parameterization of operator $\Phi$. This provides the basis to investigate optimal interpolation models and solvers for non-linear and/or non-Gaussian processes through a non-linear parameterization for operator $\Phi$. 
Here, we  benefit from the variety of neural auto-encoder architectures introduced in the deep learning literature, such as simple convolutional auto-encoders, U-Nets \cite{cicek_3d_2016}, ResNets \cite{he_deep_2016}... For such parameterization, the existence of a unique global minimum for minimization (\ref{eq: proj-based OI}) may not be guaranteed and Algorithm \ref{alg:gradient solver} will converge to a local minimum depending on the considered initial condition.

Multimodal interpolation represents another appealing extension of the proposed framework. Let us assume that some additional gap-free observation data $\mathbf{z}$ is available such that $\mathbf{z}$ is expected to partially inform state $\mathbf{x}$. We then introduce the following multimodal variational cost:
\begin{align}
\label{eq: multimodal OI}
\widehat{\mathbf{x}} =\arg \min_\mathbf{x} \lambda_1 \left \|\mathbf{y}-H_\Omega \cdot \mathbf{x}\right \|^2 & + \lambda_2 \left \| g(\mathbf{z})-h(\mathbf{x})\right \|^2 \nonumber \\ & + \lambda_3 \left \| \mathbf{x}-\Phi (\mathbf{x}) \right \|^2  
\end{align}
where $g(\cdot)$ and $h(\cdot)$ are trainable neural operators which respectively extract features from state $\mathbf{x}$ and observation $\mathbf{z}$. In this multimodal setting, $\Theta$ in (\ref{eq: training loss}) comprises the trainable parameters of operators $\Phi$, $\mathcal{G}$, $g$ and $h$. Given this reparameterization of the variational cost, we can exploit the exact same architecture for the neural solver defined by Algorithm \ref{alg:gradient solver} and the same learning setting. 

\section{Experiments}
\label{sec: experiments}

We report numerical experiments for the interpolation of 2{\sc d}+t Gaussian process (GP) for which we can compute the analytical OI solution (\ref{eq: OI solution}), as well as a real-world case-study for the reconstruction of sea surface dynamics from irregularly-sampled satellite-derived observations.

\subsection{2{\sc d}+t GP case-study}
\label{subsec: GP xp}

\paragraph{Synthetic dataset} We use the stochastic partial differential equation (SPDE) approach introduced by \citep{lindgren_2011} to generate a spatio-temporal Gaussian Process (GP). Let $\mathbf{x}$ denote the SPDE solution, we draw from the classic isotropic formulation to introduce some diffusion in the general fractional operator: 
\begin{align} 
\label{spde_diff}
\Big\lbrace{\frac{\partial{}}{\partial{t}}+\left\lbrace \kappa^2(\mathbf{s},t) - \nabla \cdot\mathbf{H}(\mathbf{s},t)\nabla \right \rbrace^{\alpha/2} \Big \rbrace} \mathbf{x}(\mathbf{s},t)=\tau \mathbf{z}(\mathbf{s},t)
\end{align}
with parameters $\kappa=0.33$ and regularization variance $\tau=1$. To ensure the GP to be smooth enough, we use a value of $\alpha=4$. Such a formulation enables to generate GPs driven by local anisotropies in space leading to non stationary spatio-temporal fields with eddy patterns. Let denote this experiment GP-DIFF2 where $\mathbf{H}$ is a 2-dimensional diffusion tensor generated by drawing from the spatial statistics literature, see e.g. \citep{fuglstad_2015a}. We introduce a generic decomposition of $\mathbf{H}(\mathbf{s},t)$ through the equation:
\begin{align*} 
\mathbf{H}=\gamma \mathbf{I}_2 + \beta \mathbf{v}(\mathbf{s})\tr\mathbf{v}(\mathbf{s})
\end{align*}
with $\gamma=1$, $\beta=25$ and $\mathbf{v}(\mathbf{s})=(v_1(\mathbf{s}),v_2(\mathbf{s}))\tr$ using a periodic formulation of its two vector fields components: it decomposes the diffusion tensor as the sum of an isotropic and anisotropic effects, the latter being described by its amplitude and magnitude. This is a valid decomposition for any symmetric positive-definite $2 \times 2$ matrix.\\
We use the Finite Difference Method (FDM) in space coupled with an Implicit Euler scheme (IES) in time to solve for the equation. Let $\mathcal{D}=[0, 100] \times [0, 100]$ be the square spatial domain of simulation and $\mathcal{T}=[0, 500]$ the temporal domain. Both spatial and temporal domains are discretized so that the simulation is made on a uniform Cartesian grid consisting of points ($x_i$, $y_j$, $t_k$) where $x_i$=$i \Delta x$, $y_j$=$j \Delta j$, $t_k$=$k \Delta t$ with $\Delta x$, $\Delta y$ and $\Delta t$ all set to one.\\
Let note that the SPDE-based formulation involves a precision-based Kalman gain $(H\tr_\Omega \mathbf{R}^{-1} H_\Omega + \mathbf{P}^{-1} )^{-1} H\tr_\Omega \mathbf{R}^{-1}$ which lies in state space while the covariance-based Kalman formulation is in the observation space. But the DFM+IES SPDE discretization lead to sparse precision matrix allowing for efficient inversion procedures. Such a formulation of the linear system in Optimal Interpolation is particularly convenient in our case to compute the regularization term of the variational cost in Eq.\ref{eq: OI}, involving the inverse of the covariance matrix. \\
To be consistent with the second dataset produced in Section \ref{xp_sat}, we sample pseudo-observations similar to along-track patterns produced by satellite datasets, with a periodic sampling leading to spatial observational rate similar to the along-track case-study. Observational noise is negligible and taken as $\mathbf{R}=\sigma^2\mathbf{I}$ with $\sigma^2=1\mathrm{E}-3$ to compute the observational term of the variational cost.

\paragraph{Benchmarked models} For the dataset GP-DIFF2, we involve a 5 timestep long spatio-temporal sequence as data assimilation window (DAW) to apply our framework and benchmark the following methods: analytical OI, as a solution of the linear system, the gradient descent OI solution, a direct CNN/UNet interpolation using a zero-filling initialization and different flavors of 4DVarnet using either a UNet trainable prior or a known precision-based prior coupled with LSTM-based solvers. As already stated in Section \label{lemma: bi-level}, we use three training loss: the mean squared error (MSE) w.r.t to the groundtruth, the MSE w.r.t to the OI analytical solution and the OI variational cost of Eq. \ref{eq: OI}. Regarding the performance metrics, we assess the quality of a model based on both OI cost value for the known SPDE precision matrix and the MSE score w.r.t to the groundtruth. We also provide the computational GPU time of all the benchmarked models on the test period. For training-free models (analytical and gradient-based OI), there is no training time. The training period goes from timestep 100 to 400 and the optimization is made on 20 epochs with Adam optimizer, with no significant improvements if trained longer. During the training procedure, we select the best model according to metrics computed over the validation period from timestep 30 to 80. Overall, the set of metrics is computed on a test period going from timestep 450 to 470.\\
Let note that by construction, the analytical OI solution is optimal regarding the OI variational cost. The same holds for its gradient based solution, when convergence is reached. Regarding the MSE, OI is unbiased with minimal variance : in other words, at a given spatio-temporal location $(\mathbf{s},t)$, its variance (which is the local MSE) is minimal. In our case, we compute the global MSE over the entire domain $\mathcal{D} \times [t_k-2,t_k+2]$ w.r.t the true state because we have only one single realization to compute this metrics. This is why OI global MSE may be outperformed by other methods.

\begin{figure}[H]
\label{gp_fig_plot}
\begin{center}
\vspace{-.5cm}
\includegraphics[width=8cm,trim=150 200 0 200,clip]{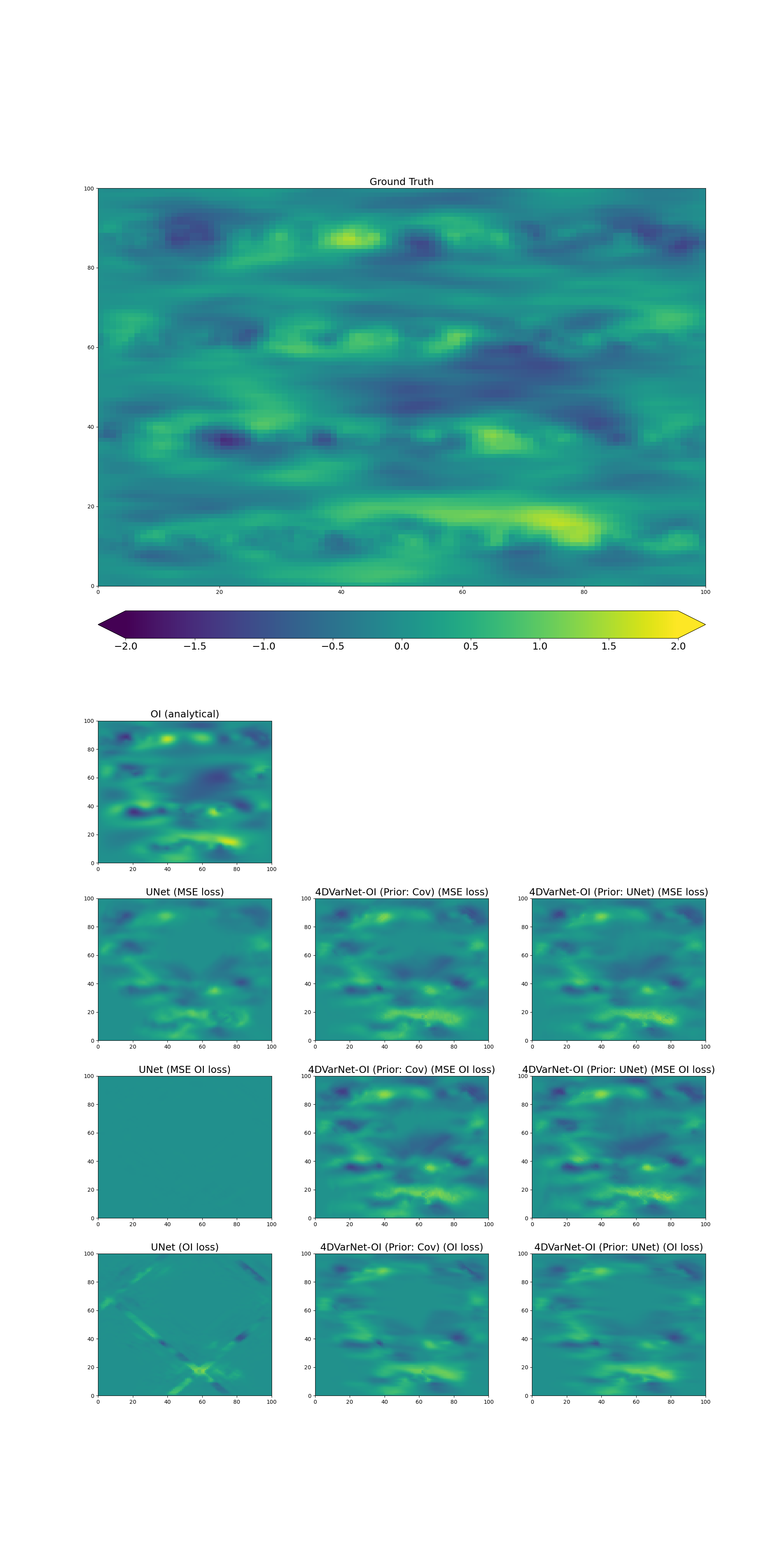}
\end{center}
\caption{SPDE-based GP spatio-temporal field (Ground Truth) and its reconstructions based on a 5 time lag assimilation window. Diffusion-based non stationary GP case}
\label{xp_GP_plot}
\end{figure}

\begin{figure}[H]
\label{gp_fig_loss}
\begin{center}
\subfloat[OI cost vs MSE]{
\hspace{.55cm} \includegraphics[width=6cm]{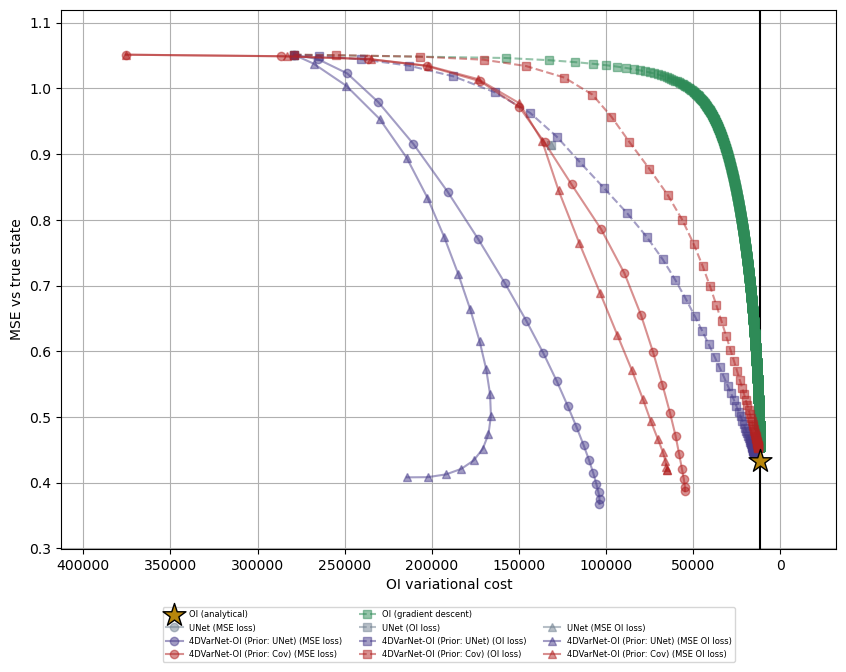}
}
\subfloat[OI cost vs number of iterations]{
\includegraphics[width=6cm]{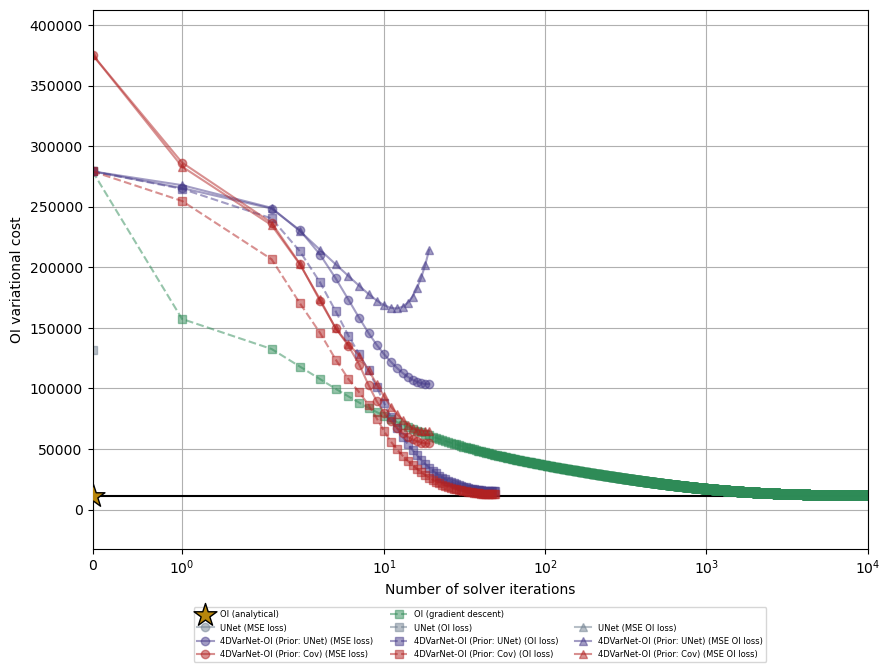}
}
\end{center}
\caption{Optimal Interpolation derived variational cost vs Mean Squared Error (MSE) loss (a) and OI variational cost vs number of iterations (b) for all the benchmarked methods at timestep 16 of the test period throughout their iterations process. For the analytical Optimal Interpolation and direct UNet neural formulations, there is no iterations, so a single point is displayed. Diffusion-based non stationary GP ($\alpha$=2) experiment}
\label{xp_GP_loss_OI_MSE}
\end{figure}

\paragraph{Results} Table \ref{table_score_GP} displays the performance evaluation for the experiment GP-DIFF2. We also provide as supplementary materials the same analysis on three other experiments denoted as GP-ISO1, GP-ISO2 and GP-DIFF1, using the same setup but with isotropic stationary formulation of the SPDE and parameter $\alpha=[2,4]$. Let note that when $\alpha=2$, the SPDE solution leads to more noisy spatio-temporal fields.\\
Figure \ref{xp_GP_plot} shows the interpolation obtained at the middle of the test period for all the benchmarked models. Because we use spatio-temporal assimilation windows of length 5, we only display the reconstructions at the center of the DAW. We also provide in Figure \ref{xp_GP_loss_OI_MSE}a) the scatterplot of the global MSE w.r.t OI variational cost throughout the iteration process, and in Figure \ref{xp_GP_loss_OI_MSE}b) the OI variational cost vs the number of iterations of the algorithm. The mapping clearly indicates that direct inversion by CNN schemes is not efficient for this reconstruction task. On the opposite, LSTM-based iterative solvers are all consistent with the optimal solution, with potential variations that can be explained by the training loss used. While using MSE w.r.t true states or OI quickly converges in a very few number of iterations, 20 typically, involving the OI variational cost as a training loss implies to increase the number of gradient steps, up to about a hundred, to reach satisfactory performance. This makes sense because using global MSE relates to supervised learning while the variational cost-based training is not. 

\begin{table*}[tb!]
    \footnotesize
    \centering
    \scalebox{.8}{
    \begin{tabular}{|p{2.cm}|p{2.cm}|p{2.cm}|p{2.cm}|p{1.5cm}|p{1.5cm}|p{2cm}|}
    \toprule
    \toprule
     \bf GP & Approach & Prior & \bf Training loss& 
     \bf MSE$_{x}$ &\bf OI-score & \bf Comp. time (mins)\\
    \toprule
    \toprule
GP-DIFF2 &                             OI &                  Covariance &             &     \textbf{2.72} & \textbf{9.8E+03} &   2.41 \\
\toprule
                           &          \multirow{3}{*}{UNet} &        \multirow{3}{*}{N/A} &    MSE loss &     3.50 & 6.10E+05 & 0.08 \\
                           &                                &                             &     OI loss &     5.99 & 1.08E+05 & 0.49 \\
                           &                                &                             & MSE OI loss &     7.38 & 4.87E+06 & 0.07 \\
\toprule
                           & \multirow{6}{*}{4DVarNet-LSTM} & \multirow{3}{*}{Covariance} &    MSE loss &     2.84 & 4.52E+04 & 2.4 \\
                           &                                &                             &     OI loss &     3.26 & 1.06E+04 & 2.68 \\
                           &                                &                             & MSE OI loss &     3.09 & 5.62E+04 &  2.33 \\
                           &                                &       \multirow{3}{*}{UNet} &    MSE loss &     2.74 & 1.04E+05 &   0.25 \\
                           &                                &                             &     OI loss &     3.17 & 1.27E+04 & 0.48 \\
                           &                                &                             & MSE OI loss &     3.03 & 2.68E+05 &  0.27 \\
    \bottomrule
    \bottomrule
    \end{tabular}
    }
    \caption{{\bf Interpolation performance for the synthetic 2D+T GP case-study:} For each benchmarked model in the isotropic case GP-DIFF1 configurations, we report the considered performance metrics for the three training loss strategies (MSE w.r.t. true states, MSE w.r.t. OI and OI variational cost)}
    \label{table_score_GP}
\end{table*}

In addition, when a similar LSTM-based solver is used, the trainable prior considerably speeds up the convergence towards the optimal solution compared to the known precision matrix-based prior. This leads to three key conclusions:
\begin{itemize}
 \item using MSE as training loss with trainable neural priors is enough for a reconstruction task and can even speed up the iterative convergence compared to the known statistical prior parametrization;
 \item the GP-based experiments demonstrates the relevance of an LSTM-based solver to speed up and accurate iterative solutions of minimization problems;
 \item looking for an optimal solution within the bi-level neural optimization of prior and solver may lead to deviate from the original variational cost to minimize. 
\end{itemize} 

\subsection{Satellite altimetry dataset}
\label{xp_sat}

\paragraph{Experimental setting} We also apply our neural OI scheme to a real-world dataset, namely the interpolaton of sea surface height (SSH) fields from irregularly-sampled satellite altimetry observations. The SSH relates to sea surface dynamics \citep{LeGuillou_2020} and satellite altimetry data are characterized by an average missing data rate above 90\%. We exploit the experimental setting defined in \citep{LeGuillou_2020} \footnote{SSH Mapping Data Challenge 2020a: \url{https://github.com/ocean-data-challenges/2020a_SSH_mapping_NATL60}}. It relies on a groundtruthed dataset given by the simulation of realistic satellite altimetry observations from numerical ocean simulations. Overall, this dataset refers to 2{\sc d}+t states for a $10^\circ \times 10^\circ$ domain with 1/20$^\circ$ resolution corresponding to a small area in the Western part of the Gulf Stream.  

Regarding the evaluation framework, we refer the reader to SSH mapping data challenge above mentioned for a detailed presentation of the datasets and evaluation metrics. The latter comprise the MSE w.r.t the Ground Truth, the minimal spatial and temporal scales resolved, and we also look for the relative gain w.r.t DUACS OI for SSH and its gradient. 

For learning-based approaches, the training dataset spans from mid-February 2013 to October 2013, while the validation period refers to January 2013. All methods are tested on the test period from  October 22, 2012 to December 2, 2012.

\paragraph{Benchmarked models:}  For benchmarking purposes, we consider the approaches reported in 
\citep{LeGuillou_2020}, namely: the operational baseline (DUACS) based on an optimal interpolation, multi-scale OI scheme MIOST \citep{Ardhuin_2020} and model-driven interpolation schems BFN \citep{LeGuillou_2020} and DYMOST \citep{Ubelmann_2016, Ballarotta_2020}. We also include a state-of-the-art UNet architecture to train a direct inversion scheme \cite{cicek_3d_2016}. For all neural schemes, we consider 29-day space-time sequences to account for time scales considered in state-of-the-art OI schemes. Regarding the parameterization of our framework, we consider a bilinear residual architecture for prior $\Phi$, a classic UNet flavor as well as a simple linear convolutional prior. Similarly to the GP case-study, we use a 2{\sc d} convolutional LSTM cell with 150-dimensional hidden states. Besides the interpolation scheme using only altimetry data, we also implement a multimodal version of our interpolation framework. It uses sea surface temperature (SST) field as complementary gap-free observations. SST fields are widely acknowledge to convey information on sea surface dynamics though the derivation of an explicit relationship between SSH and SST fields remain a challenge, except for specific dynamical regimes \citep{isern-fontanet_potential_2006}. Our multimodal extension exploits simple convnets for the parameterization of operators $g(\cdot)$ and $h(\cdot)$ in Eq.\ref{eq: multimodal OI}.

\begin{figure}[H]
\begin{center}
\includegraphics[width=9cm,trim=50 0 0 0,clip]{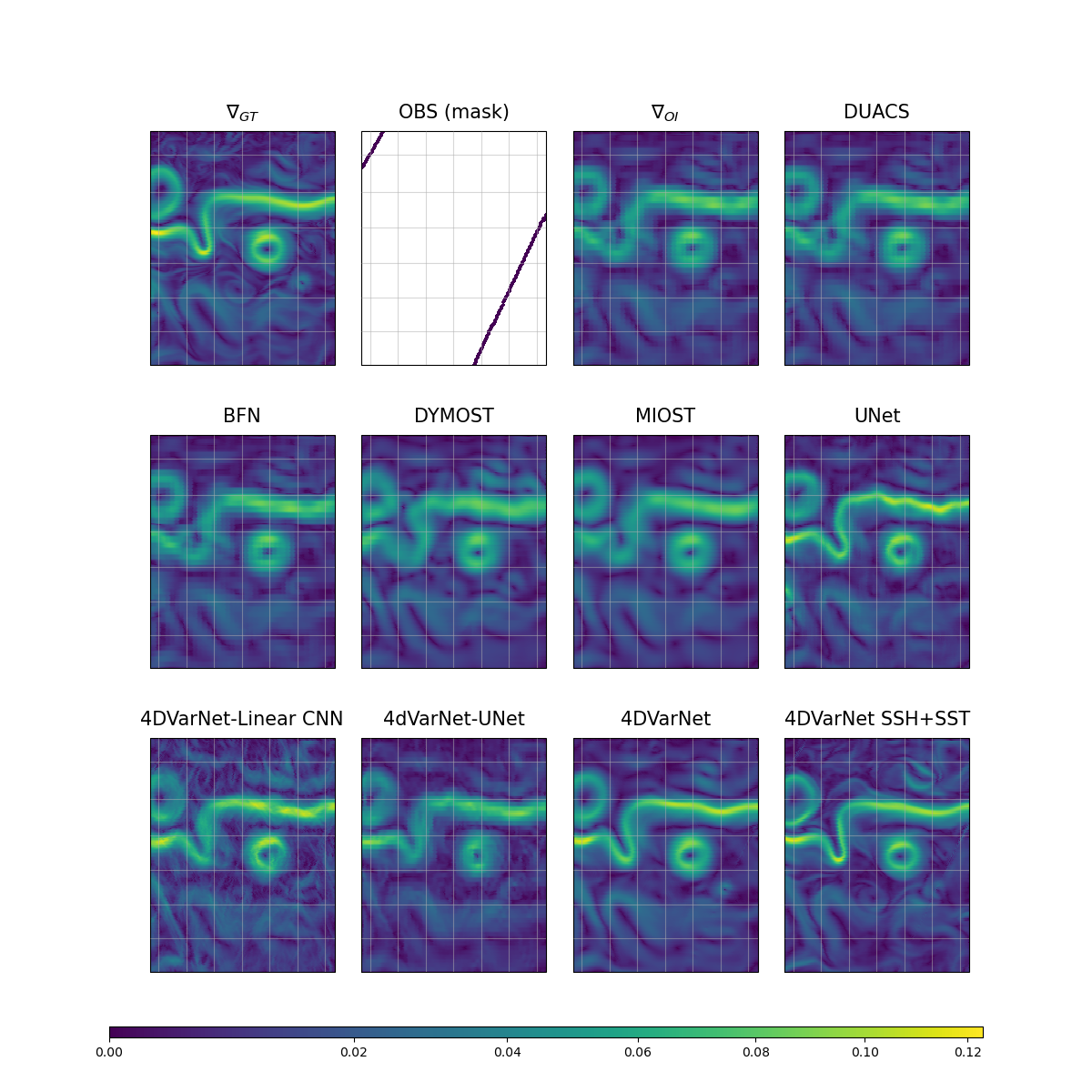}
\end{center}
\caption{Gradient SSH reconstructions (2012-11-05) for all benchmarked models based on 4 along-track nadirs pseudo-observations in the Gulf Stream domain. Trom top left to bottom right: Ground Truth, Observations, naive OI, DUACS, BFN, DYMOST, MIOST, UNet, 4DVarNet with a linear CNN-based prior, 4DVarNet with UNet prior, 4DVarNet with BiLin-Res prior and multimodal 4DVarNet with BiLin-Res prior embedding SSH-SST synergies in the variational cost}
\label{xp_SSH}
\end{figure}

\begin{table*}[tb]
    \footnotesize
    \centering
    \scalebox{.7}{
    \begin{tabular}{|p{3.cm}|p{2cm}|p{1.5cm}|  p{1.5cm}|p{1.5cm}|p{1.5cm}|p{1.5cm}|}
    \toprule
    \toprule
     \bf Approach & Prior &
     \bf MSE & $\lambda_x$ (degree)& $\lambda_t$ (days)&
     $\tau_{SSH}$ (\%) & $\tau_{\nabla SSH}$ (\%)\\
    \toprule
    \toprule
    DUACS & - & 0.92 & 1.42 & 12.13 & - & - \\
    BFN & - & 0.92 & 1.23 & 10.82 & 7.93& 23.69\\
    DYMOST & - & 0.91 & 1.36 & 11.91 & -10.88 & 0.38\\
    MIOST & - & 0.93 & 1.35 &10.41 & 25.63 & 11.16 \\
    \toprule
    UNet & - & 0.924 & 1.25  & 11.33  & 20.13 & 26.16 \\
    \toprule
    \multirow{3}{*}{4DVarNet-LSTM} & Linear CNN & 0.89 & 1.46 & 12.63 & -84.14 & -10.24 \\ 
     & UNet & 0.89 & 1.4& 12.45&0.24 & 0.01 \\
     & BiLin-Res & \textbf{0.94} & \textbf{1.17} & \textbf{6.86} & \textbf{54.79} & \textbf{55.14}  \\
    \toprule
        \toprule
    \multicolumn{7}{|c|}{\bf Multimodal interpolation models (SSH+SST)}\\
        \toprule
    UNet & - & 0.55 & 2.36 & 35.72 & -2741.29& -355.24 \\
    4DVarNet-LSTM &  BiLin-Res & \textbf{0.96} & \textbf{0.66}  & \textbf{2.97} & \textbf{79.80}& \textbf{75.71}\\
    \bottomrule
    \bottomrule
    \end{tabular}
    }
    \caption{{\bf Interpolation performance for the satellite altimetry case-study:} For each benchmarked models, we report the considered performance metrics averaged on the test period when learning-based methods are trained on the MSE loss (true states and its gradient)}
    \label{tab_score_SSH} 
\end{table*}

\paragraph{Results} Figure \ref{xp_SSH} displays the  reconstructions of the SSH field and the corresponding gradients on 2012-11-05 for all the benchmarked models. It clearly stresses how our scheme improves the reconstruction when considering a non-linear prior. Especially, we greatly sharpen the gradient along the main meander of the Gulf Stream compared with other interpolation schemes. Oceanic eddies are also better retrieved. Table \ref{tab_score_SSH} further highlights the performance gain of the proposed scheme. The relative gain is greater than 50\% compared to the operational satellite altimetry processing. We outperform by more than 20\% in terms of relative gain to the baseline MIOST and UNet schemes, which are the second best interpolation schemes. Interestingly, our scheme is the only one to retrieve time scales below 10 days when considering only altimetry data.

As stressed by last line and map of Table \ref{tab_score_SSH} and Figure \ref{xp_SSH}, the multimodal version of the proposed interpolation scheme further improves the interpolation performance. Our trainable OI solver learns how to extract fine-scale features from SST fields to best reconstruct the fine-scale structure of SSH fields. The relevance of this multimodal setting is visually-explicit and results in a significant improvement of all performance metrics.

\section{Conclusion}
\label{sec: conclusion}

This paper addresses the end-to-end learning of neural schemes for optimal interpolation. We extend the neural scheme introduced in  \citep{Fablet_2021} for data assimilation to optimal interpolation with theoretical guarantees so that the considered trainable solvers asymptotically converge towards the analytical OI solution. Numerical experiments for synthetic GP datasets and ocean remote sensing case-studies support the relevance of the proposed scheme compared with state-of-the-art interpolation techniques, when dealing with very large missing data rates. 

\paragraph{Trainable and scalable OI solvers:} The computation of the analytical OI solution is challenging when dealing with high-dimensional states. Whereas classic gradient-based iterative methods may suffer from a relatively low convergence rate, our experiments support the relevance of the proposed trainable solvers to speed up the convergence and reach good interpolation performance with only 10 to 100 gradient steps. Importantly, the convolutional architecture of the trainable solver also guarantees their scalability and a linear complexity with respect to the size of the spatial domain as well as the number of observations.

\paragraph{End-to-end learning for optimal interpolation:} Our GP experiments highlight the relevance of the bi-level formulation of the OI problem. It allows us to explore different configurations for the inner variational cost and the outer training loss. We can guarantee the convergence to the OI solution, when considering the OI variational cost as the inner cost. But, we may also greatly speed up the interpolation time, when considering a Unet-based parameterization of the inner cost and the interpolation error as the outer performance metrics. The latter strategy greatly simplifies the application of the proposed framework to real datasets, where the underlying covariance model is not knwown and/or a Gaussian process approcimation does not apply. As illustrated for our application to ocean remote sensing data, the proposed framework greatly outperforms both geostatistical, learning-based and model-driven techniques, especially when benefitting from additional multimodal observations.

\paragraph{Opened questions} Whereas in the GP case, we know the variational OI cost to be the optimal variational formulation to solve the interpolation in terms of interpolation performance. No such theoretical result in most non-Gaussian/non-linear cases. The proposed end-to-end learning framework provides new means to explore the reduction of estimation biases in Bayesian setting as well as to investigate. Especially, our  experiments on ocean remote sensing data suggest that the prior term in the inner variational formulation shall be adapted to the observation configuration rather than considering generic plug-and-play priors.


\end{document}